\documentclass[showpacs,twocolumn,floats,superscriptaddress]{revtex4}
\usepackage{amssymb}
\usepackage{graphicx}
\usepackage{amsmath}
\usepackage{bm}

\newcommand{\bq}{\begin{equation}}
\newcommand{\ee}{\end{equation}}
\newcommand{\fr}[2]{\frac{#1}{#2}}
\newcommand{\eps}{\varepsilon}
\begin{document}

\title{Charge accumulation at the boundaries of a graphene
strip induced by a gate voltage: Electrostatic approach
}
\author{P.G. Silvestrov}
\affiliation{Theoretische Physik III,
Ruhr-Universit{\"a}t Bochum, 44780 Bochum, Germany}
\author{K.B. Efetov }
\affiliation{Theoretische Physik III,
Ruhr-Universit{\"a}t Bochum, 44780 Bochum, Germany}
\affiliation{L. D. Landau Institute for Theoretical
Physics, 117940 Moscow, Russia}
\date{\today }

\begin{abstract}
Distribution of charge induced by a gate voltage in a graphene
strip is investigated. We calculate analytically the charge
profile and demonstrate a strong(macroscopic) charge accumulation
along the boundaries of a micrometers-wide strip. This charge
inhomogeneity is especially important in the quantum Hall regime
where we predict the doubling of the number of edge states and
coexistence of two different types of such states. Applications to
graphene-based nanoelectronics are discussed.
\end{abstract}

 \maketitle

The new material graphene, a monolayer of carbon atoms with
honeycomb lattice structure, is attracting a lot of interest since
2005 when the first transport measurements in this material have
been reported~\cite{Novoselov04,Novosel05,Zhang05}. The interest
in 2D electron gases in graphene originates from the Dirac-like
spectrum of the low-energy quasiparticles~\cite{Fse}. Several
prominent phenomena have been investigated in this
\textquotedblleft relativistic\textquotedblright\ system both
experimentally and theoretically, including quantum Hall effect
(QH) \cite{Gsin05,Abanin06}, weak localization and other effects
of disorder~\cite{McCann06,AlEf06}, superconducting proximity
effects~\cite{Beenakker06,Titov06Sept,GDSS07,VanDerSypen07}, etc.

In the experiments~\cite{Novoselov04,Novosel05,Zhang05},
mechanically exfoliated graphene samples were separated from the
metallic gate by a $b\approx 0.3\mu m$ wide insulating
layer~($SiO_{2}$). The width of the insulator is dictated by the
necessity to identify optically the single-layer graphene. In the
undoped graphene (half filling), the charge of the conduction
electrons is compensated by the charge of the carbon ions forming
the lattice. By applying a large ($V_{g}\lesssim 100\mathrm{V}$)
voltage $V_{g}$ to the lower gate one induces a considerable
($n_{e}/V_{g}\approx 7.2\
10^{10}\mathrm{cm}^{-2}/\mathrm{V}$~\cite{Novoselov04})
uncompensated charge $e \times n_{e}$ in the graphene plane. This
extra charge is screened by \textquotedblleft image
charges\textquotedblright\ induced in the metallic gate. However,
since the images are located $0.6\mu m$ below graphene, such a
screening becomes effective only in the central region of several
microns large graphene samples. As a result, the charge
distribution cannot be homogeneous.

\begin{figure}[tbp]
\includegraphics[width=7.cm]{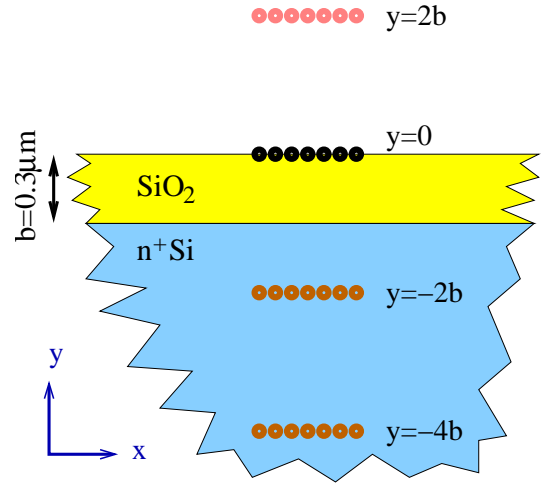} 
\caption{Considered experimental setup. Charges (doped graphene
strip) are placed on the surface ($y=0$) of $0.3\mu m$ thick
insulating layer ($SiO_2$, dielectric constant $\varepsilon =3.9$)
above the metallic ($n^+Si$) gate. Image charges(charged strips)
are shown above and below the graphene plane. } \vspace{-.3cm}
\end{figure}

In this paper, we calculate analytically the charge distribution
in the graphene strip and demonstrate a strong increase of the
charge density near the strip edges (numerically, a charge
accumulation near the edges has been seen in Ref.~\cite{Brey07}).
For a gate voltage of $\approx 10{\rm V}$ the distance between the
excess electrons in the sheet is of order $\sim 10nm$. This means
that for the $0.1\div 1\mu m $ wide strips one may speak of a
continuous charge distribution and determine the latter minimizing
the electrostatic energy of the electrons. In semiconductor
heterostructures the electron redistribution has been discussed in
the context of compressible/incompressible QH stripes
formation~\cite{Chklovskii92}. However, in that case electrons
were confined by a smooth potential, which resulted in a
continuous charge density profile at the edge. As we will see, at
the sharp graphene edge the charge develops a $1/\sqrt{x}$
singularity.

The charge inhomogeneity discussed in the present paper develops
at the scale $\sim 0.1$ micrometer. So this is a macroscopic
effect that should have clear experimental consequences. The
distribution of classical excess charges found below is valid for
any metallic strip. However, only in graphene the excess charge
density coincides with the carrier density and determines directly
the Fermi momentum, $p_F\propto \sqrt{n_e}$. The charge
accumulation at the graphene boundaries is especially important
for the quantum Hall effect where we predict coexistence of two
types of edge states~\cite{Chklovskii92,Halperin82,Chamon94}. The
strong dependence of the charge density on the strip width may
have interesting nanoelectronic applications, discused at the end
of the paper.

Electrostatic potential created by the charge located on the
surface of the insulator, both above and inside, 
is given by~\cite{LandauMedia}
 \begin{equation}
\phi =\frac{2}{1+\varepsilon }\frac{e}{|{\bm R}|},  \label{chargebare}
 \end{equation}
where the dielectric constant $\varepsilon _{SiO_{2}}=3.9$. In
order to describe the potential created by the charge placed on a
dielectric layer ($-b<y<0$) with a metallic gate attached
underneath ($y<-b$), one has to consider potentials of a string of
image charges, as illustrated by Fig.~1. [We reserve the
coordinates $x$ and $z$ for the graphene plane, or the surface of
the insulator.]
Two different expressions describe now the potential inside the
insulator, at $-b<y<0$,
 \begin{equation}
\phi =\sum_{n=0}^{\infty }\frac{2e\xi ^{n}}{1+\varepsilon }\left(
\fr{1}{|{\bm R} -2n{\bm b}|}-\fr{1}{|{\bm R}+2n{\bm b}+2{\bm
b}|}\right) , \label{phibelow}
 \end{equation}
and above it, at $y>0$,
 \bq\label{phiabove}
\phi =\frac{2}{1+\varepsilon }\frac{e}{|{\bm
R}|}-\sum_{n=1}^{\infty }\frac{4\varepsilon \xi
^{n}}{1-\varepsilon ^{2}}\frac{e}{|{\bm R}+2n{\bm b}|}.
 \ee
Here $\xi =(1-\varepsilon )/(1+\varepsilon )$, and the vector
${\bm b}$, $|{\bm b}|=b$ is directed along the $y$-axis,
perpendicular to graphene plane. One may easily show that the
potential Eqs.~(\ref{phibelow},\ref{phiabove}) satisfy the
boundary conditions $\phi_1=\phi_2$ and
$\eps_1\partial\phi_1/\partial y=\eps_2\partial\phi_2/\partial y$
at the surface of the insulator, $y=0$, and Eq.~(\ref{phibelow})
gives $\phi={\rm const}=0$ at the metallic surface, $y=-b$. Fig. 2
shows the in-plane potential found from these formulas. This
potential describes the electron interaction in mechanically
exfoliated graphene.

\begin{figure}[tbp]
\includegraphics[width=7.cm]{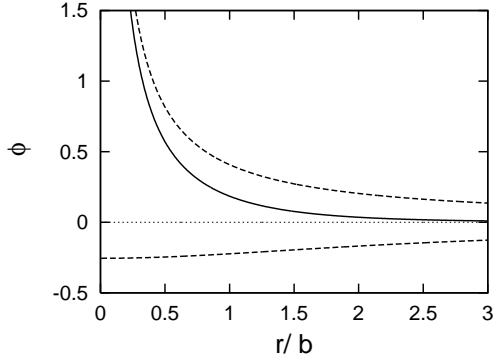} 
\caption{Screening of the electron potential on the surface of the
insulator ($\varepsilon =3.9$), as described by
Eqs.~(\ref{phibelow},\ref{phiabove}). Solid line shows the
screened potential; upper dashed line is the potential of the
charge on the surface without image charges included,
Eq.~(\ref{chargebare}), and the lower dashed line stands for the
separate contribution due to image charges. We measure all
distances in units of the width of insulating layer, $b=0.3\mu m$,
which is fixed in the experiment. For the in-plane distance
$r=\sqrt{x^2+z^2}\gg b$ we have $\protect\phi \approx {2}{b^{2}e}
/ \protect\varepsilon ^{2}{r^{3}}$.} \vspace{-.3cm}
\end{figure}

Let us consider now a narrow graphene strip with the width $2a$,
such that $a\ll b$, directed in the $x,z$ plane along $z$-axis.
Since the charge is distributed uniformly along the strip, the
potential both inside and above the insulating $SiO_{2}$ may be
obtained from the real part of a holomorphic function
$w(\mathcal{Z})$ as $\phi =\mathrm{Re}w(\mathcal{Z})$,
$\mathcal{Z}=x+iy$, $\Delta w(\mathcal{Z})\equiv 0$. In
particular, the function
 \begin{equation}
\phi _{0}=\frac{4\sigma }{1+\varepsilon }L(x+iy)\ ,\
L(\mathcal{Z})=\mathrm{Re}\ln
\frac{\mathcal{Z}-\sqrt{\mathcal{Z}^{2}-a^{2}}}{a},
\label{phiwithout}
 \end{equation}
is a solution of the Poisson equation $\Delta \phi _{0}=-4\pi \rho _{0}$
with the charge density
 \begin{equation}
\rho _{0}=\frac{\sigma }{\pi }\frac{\delta (y)}{\sqrt{a^{2}-x^{2}}},
\label{rhobare}
 \end{equation}
where $\delta (y)$ is the delta-function and $\sigma $ is the
charge per unit length of the strip. The factor $2/(1+\varepsilon
)$ in Eqs.~(\ref{chargebare},\ref{phiwithout}) accounts for the
polarization of the dielectric substrate. The function $\rho _{0}$
(\ref{rhobare}) is the  equilibrium charge distribution, since the
potential $\phi _{0}$, is constant on the strip, $\phi \equiv 0$
at $-a<x<a,y=0$~(\ref{phiwithout}). The inverse square root edge
singularity in Eq.~(\ref{rhobare}), $\rho \sim 1/\sqrt{x-a}$,
{drastically differs from} the square root density profile $\rho
\sim \sqrt{x}$ at the soft wall edge in the conventional
heterostructures~\cite{Chklovskii92}.

Straightforward generalization of Eq.~(\ref{phibelow}) gives the
potential inside the insulating layer sandwiched between the
metalic gate and narrow ($a\ll b$) graphene strip
 \begin{equation}
\phi =\sum_{n=0}^{\infty }\frac{4\sigma \xi ^{n}}{1+\varepsilon
}[L(\mathcal{Z}-2inb)-L(\mathcal{Z}+2i(n+1)b)],
\label{phistripbelow}
 \end{equation}

Equations (\ref{rhobare}) and (\ref{phistripbelow}) are the
central result of this paper describing the charge and potential
distributions in the narrow mechanically exfoliated graphene
strip. Below we show that these results remain quantitatively
accurate even for sufficiently wide strips, when $a\approx b$.

After the image charges are added the potential on a strip
acquires a small, $\sim (a/b)^{2}$, coordinate dependent
correction. From Eq.~(\ref{phistripbelow}) at $y=0$, $-a<x<a$ we
find
 \begin{equation}
\phi (x,0)=\frac{4\sigma }{\varepsilon +1}\left[ \ln
\frac{4b}{a}+C_{0}+\frac{2x^{2}+a^{2}}{2b^{2}}C_{2}+\cdots \right]
,  \label{phistrip}
 \end{equation}
where $C_{0}=\sum_{n=2}^{\infty }\frac{2\varepsilon
}{1-\varepsilon }\xi ^{n}\ln n$, $C_{2}=\sum_{n=1}^{\infty
}\frac{\varepsilon }{1-\varepsilon } \xi ^{n}{n^{-2}}$. For
$\varepsilon =3.9$ we found $C_{0}\approx -0.31$ and $C_{2}\approx
0.175$.

\begin{figure}[tbp]
\includegraphics[width=7.cm]{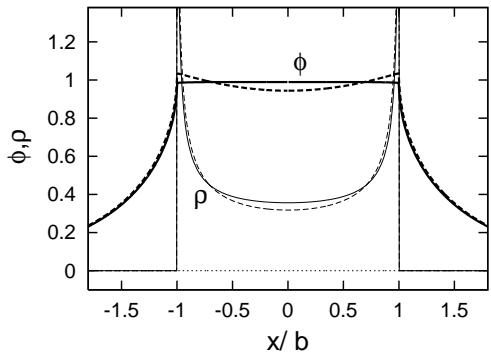} 
\caption{Potential (thick lines) and density (thin lines) across
the strip of width $2a=2b=0.6\protect\mu m$. The dashed lines show
the curves for the density $\protect\rho_0$,
Eq.~(\protect\ref{rhobare}) and the solid ones - for the density
$\protect\rho=\protect\rho_0-0.12\protect\rho_2$,
Eq.~(\ref{phistripAll}). At the plateau one has
$\protect\phi(x)\approx 0.99\protect\sigma$. A gate voltage
$V_g=100{\rm V}$ creates in such a strip an averaged electron
density $\langle n\rangle=11.6\times 10^{12}cm^{-2}$ and the
minimal density $n(0)=8.25\times 10^{12}cm^{-2}$, while an
infinite graphene plane gives $n_{\infty}=7.2\times
10^{12}cm^{-2}$~\cite{Novoselov04}. Semiclassical approximation
used for the density calculation breaks at the distance $\delta
x/b> 0.05({\rm V}/V_g)^{2/3}$ from the boundary.} \vspace{-.3cm}
\end{figure}

The coordinate dependence of the potential
Eqs.~(\ref{phistripbelow},\ref{phistrip}) on the {\it metallic}
strip should be compensated by a proper charge redistribution.
Since $\phi (x)$, Eq. (\ref{phistrip}), increases towards the
edges, one should transfer some charges from the boundaries to the
strip center. To find the equilibrium distribution for finite
$a/b$ we add a series of \textquotedblleft
multipole\textquotedblright\ corrections to the potential $\phi
_{0}$, Eq.~(\ref{phiwithout})~\cite{Raikh06},
 \begin{equation}
\phi =\phi _{0}+\sum_{n=1}\alpha _{n}\phi _{2n}\ ,\ \phi
_{j}=\mathrm{Re} \frac{2\sigma
(\mathcal{Z}-\sqrt{\mathcal{Z}^{2}-a^{2}})^{j}}{(1+\varepsilon
)a^{j}}.  \label{phistripAll}
 \end{equation}
The same corrections should be added to all image strip potentials
in Eq.~(\ref{phistripbelow}). Corresponding corrections to the
charge density, Eq.~(\ref{rhobare}), are
 \begin{equation}
\rho =\rho _{0}+\sum_{n=1}\alpha _{n}\rho _{2n}\ ,\ \rho
_{j}=-\frac{1+\varepsilon }{4\pi }\delta (y)\left. \frac{d\phi
_{j}}{dy}\right\vert _{y=+0}.  \label{rhostripAll}
 \end{equation}
For example, $\rho _{2}=(2x^{2}/a^{2}-1)\rho _{0}$. Still the
singularity $\rho \sim 1/\sqrt{x-a}$ at the edge is generic for
any strip width.

To compensate the $\sim x^{2}$ term in Eq.~(\ref{phistrip}) it is
enough to consider the first correction only,
$\alpha_{2}=-0.175(a/b)^{2}$. This allows us to approximate the
equilibrium distribution $\phi (x)=\mathrm{const}$ with the
accuracy better than $0.2\%$ for $a<0.5b$ (strip width $2a<0.3\mu
m$). A simple formula (both $V_g$ and $\sigma$ have dimensionality
{\it charge/distance})
 \begin{equation}
V_{g}=\sigma \lbrack 0.82\ln (b/a)+0.88+0.29(a/b)^{2}]
\label{gatecharge}
 \end{equation}
relates in this case the linear charge density $\sigma $ in the
narrow strip to the applied gate voltage.

An appropriate fit with only two parameters
$\alpha_{2},\alpha_{4}\neq 0$ allows us to reach $\phi (x)\approx
\mathrm{const}$ on the strip with the accuracy $\sim 0.5\%$ even
for $a=5b$ (strip width $2a=3\mu m$). Remarkably, even for such a
wide strip the amplitude of the $1/\sqrt{x}$ singularity is
reduced only by a factor $0.55$ compared to the simple formula,
Eq.~(\ref{rhobare}). Fig.~3 shows results of the single parameter
fit, $\alpha_{2}\neq 0$, for $a=b$.

\begin{figure}[tbp]
\includegraphics[width=7.cm]{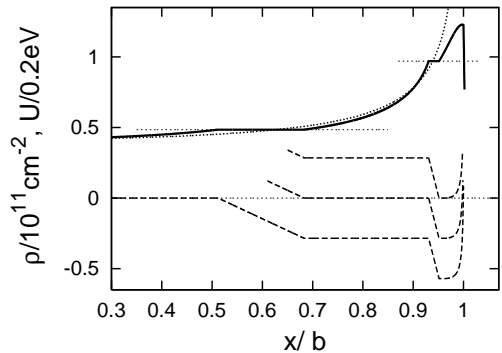} 
\caption{Schematic charge distribution $\rho(x)$ in units of
$10^{11}cm^{-2}$ in the QH regime for a strip width $2a=2b=0.6\mu
m$, $B=10T$, $V_g=5V$ (thick solid line). The $\sim 1/\sqrt{x}$
increase of density at the edge is stopped at $\Delta x\sim l_B=
\sqrt{\hbar c/eB}\approx 8nm$. Short-dashed line shows the
electrostatic solution, the same as in Fig.~2. We assume
valley-degenerate Landau levels, and choose the Zeeman splitting
$E_{\mathrm{Zeeman}}=0.25E_0$~\protect\cite{Abanin06}. Lower
dashed curves show the effective potential for different Landau
levels $U_{\mathrm{eff}}(x)=U(x)+\protect\sqrt{N}E_0\pm
E_{\mathrm{Zeeman}}$ (in units of $0.2eV$). All electronic states
with $U_{\mathrm{eff}}<0(>0)$ are occupied(empty). Regions with
$U_{\mathrm{eff}}=0 $ correspond to partially occupied Landau
levels (compressible stripes). The figure shows coexistence of two
types of edge states: Compressible stripes in the center and usual
noninteracting edge states at the borders.} \vspace{-.3cm}
\end{figure}

The classical charges equilibration condition, applicable for any
metallic strip, allows to find the inhomogeneous electron density,
Eqs.~(\ref{rhobare},\ref{rhostripAll}), but leads to the constant
potential in plane. A nontrivial graphene-specific potential
profile across the strip appears due to quantum effects. Quantum
dynamics of electrons in graphene is described by the Dirac
equation
 \begin{equation}
[ v_{F}(\tau _{x}p_{x}\pm \tau _{y}p_{y})+U(x)]\psi _{\pm
}=\varepsilon \psi _{\pm },  \label{Dirac}
 \end{equation}
where $\tau _{x,y}$ are the Pauli matrices interchanging the
sublattice index on the honeycomb lattice. [Strictly speaking we
should write here $(\tau _{x}p_{x}\pm \tau _{y}p_{z})$, since we
use coordinates $x,z$ for the graphene plane, not $x,y$ used
usually in the literature.] The two signs $\pm $ correspond to two
valleys in graphene and $\psi _{\pm }$ are envelope functions.
Solutions of Eq.~(\ref{Dirac}) are double degenerate due to spin
and $v_{F}\approx 10^{8}cm/s$. The Pauli principle prevents all
uncompensated electrons in the strip from having the same zero
momentum, $p=0$, as was assumed in the electrostatic solution,
Eq.~(\ref{rhobare}). To account for the coordinate dependent
electron density we introduced in Eq.~(\ref{Dirac}) a potential
$U(x)<0$, while keeping the zero Fermi energy, $E_{F}\equiv 0$.
For a large charge density, the potential $U(x)$ varies slowly on
the scale of the wave length $\lambda $, which allows us to
introduce the local Fermi momentum $p_{F}=\hbar /\lambda
_{F}=|U(x)|/v_{F}$. The density of electrons can be found in the
Thomas-Fermi approximation as
 \begin{equation}
n_{e}=4\int_{|p|<p_{F}}\frac{d^{2}p}{(2\pi \hbar
)^{2}}=\frac{1}{\pi }\left( \frac{U(x)}{\hbar v_{F}}\right) ^{2}.
\label{e1}
 \end{equation}
The 2-dimensional density of electrons $n_e$ is related to the
3-dimensional charge density used in
Eqs.~(\ref{rhobare},\ref{rhostripAll}) as $\rho= e n_e\delta(y)$.
Thus for narrow strip, $a\ll b$, we find from
Eqs.~(\ref{rhobare},\ref{e1})
 \begin{equation}
U(x)=-\hbar v_{F}\sqrt{\sigma/e}(a^2-x^2)^{-1/4}. \label{VThF}
 \end{equation}
For the gate voltage $V_g=100{\rm V}$ and the strip width
$2a=0.6\mu m$ we estimate $U(0)=-0.335e\! \mathrm{V}$. This
quantum ($U(x)\propto\hbar$) correction to the electrostatic
potential on the strip describes locally the position of the Dirac
crossing point with respect to Fermi energy.

The semiclassical approximation used here is justified provided
$|d\lambda_{F}/dx|\ll 1$. Thus we may use Eqs.~(\ref{rhobare},
\ref{VThF}) only at distances $\delta x>a^{1/3}(e/\sigma )^{2/3}$
from the strip edge. At $\delta x \sim a^{1/3}(e/\sigma )^{2/3}$
the singular increase of both the density Eq.~(\ref{rhobare}) and
the potential Eq.~(\ref{VThF}) is stopped in a way dependent on
the details of the graphene edge. In particular, the maximal value
of the density is $n_{\rm max}={\rm const} \times (\sigma^2/e^2
a)^{2/3}$ with ${\rm const}\sim 1$ depending on the type of the
edge.

Experimentally, the conductivity $\Sigma$ of graphene increases
linearly with the gate
voltage~\cite{Novoselov04,Novosel05,Zhang05}. This implies $\Sigma
\propto v_{F}p_{F}\propto \sqrt{\rho }$. The increase of the
carrier density near the edges of the strip should lead to an
inhomogeneous current density distribution $j(x)\propto \sqrt{\rho
(x)}$. Presence of a (moderately strong) disorder should not
change the density distribution in the strip.

The non-monotonic charge distribution across the graphene
strip~\cite{cleaved} should be especially important in the QH
regime when the electron transport is due to existence of edge
states~\cite{Halperin82}. The charge density in this case is
roughly given again by Eq.~(\ref{rhobare}) with the $1/\sqrt{x}$
edge singularity smoothed at the distances $\sim l_B=\sqrt{\hbar
c/eB}$. The number of occupied Landau levels as a function of the
transverse coordinate $x$ first increases almost abruptly (at the
length $\sim l_{B}$) at the strip edge and then decreases towards
the strip center (length scale $\sim a$). This leads to formation
of a double set of QH edge states, having a very different
microscopic nature (see Fig.~4). At the graphene edge one may
effectively neglect the electron interaction and consider the QH
edge states formed by the last occupied electron state on the
$n$-th branch of solutions of single-particle Dirac equation
$E_{n}(x_{0})$ (the $n$-th Landau
level)~\cite{Halperin82,Chamon94}. This is an adequate description
of the outer edge states in graphene~\cite{Abanin06}. Away from
the boundaries the electron repulsion transforms the narrow ($\sim
l_{B}$) edge states into the wide compressible stripes with a
macroscopic number of electrons having similar energies
$U_{\mathrm{eff}}(x)=\mathrm{const}.$ These compressible stripes
alternate with the incompressible ones having a constant electron
density $n_{e}(x)=\mathrm{const}$~\cite{Chklovskii92}.

\begin{figure}[tbp]
\includegraphics[width=5.7cm]{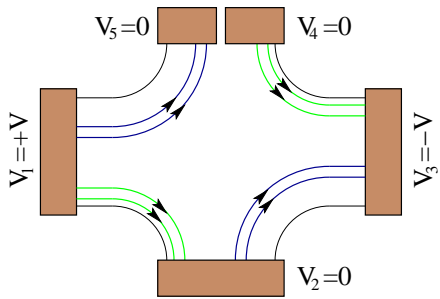} 
\caption{Five-terminal device for measuring separately the Hall
currents in inner and outer edge states. Only the current carrying
channels are shown. The two currents cancel each other in the gate
$2$, while the split gates $4$ and $5$ measure each the current
due to several inner or outer states.} \vspace{-.3cm}
\end{figure}

The energy of the $N$-th Landau level in graphene is~\cite{Gsin05}
 \begin{equation}
E\approx \sqrt{N}E_{0}\ ,\ E_{0}=\hbar v_{F}\sqrt{{2eB}/{\hbar c}}.
 \end{equation}
There is a series of such levels for each spin and valley component. So, we
find the number of occupied Landau levels at a given point in a strip
 \begin{equation}\label{Nsmooth}
N(x)=\left( {U(x)}/{E_{0}}\right) ^{2}=n_e hc/4eB.
 \end{equation}
This formula shows the smoothed number of occupied Landau levels
for $N(x)\gg 1$. Going beyond this approximation reveals the
compressible and incompressible striped QH states shown in Fig.~4.
The picture is schematic since the details of
compressible/incompressible stripes are not described by the
smooth Eq.~(\ref{Nsmooth}). [Still this may be done
electrostatically, see Ref.~\cite{Chklovskii92}, in the regions
there $dU/dx\ll E_0/l_B$.] The form of the density close to the
edge, $\delta x\sim l_B$, as well as the physics of outer edge
states~\cite{Abanin06,Fertig06}, depends on the form of graphene
edge. Nevertheless we may say that the lower gate voltage
$V_{g}\approx 5V$ should be sufficient to create several edge
states of both kinds for the strip width $ 2a\approx 0.6\mu m$.

Each QH edge channel supports the electric current flowing only in
one direction. For the appropriate sign of the bias voltage $V$,
the channel may carry the current $j=e^2V/h$. Since electrons
belonging to inner and outer QH edge channels drift in opposite
directions, corresponding currents have a tendency to compensate
each other. In Fig.~5 we suggest a simple five-terminal device,
which would allow to measure separately the currents carried by
the outer and inner channels~\cite{endnoteQH}. The QH effect in
such a setup would be seen at parametrically smaller values of the
gate voltage than in the existing
experiments~\cite{Novosel05,Zhang05}.

\begin{figure}[tbp]
\includegraphics[width=7.5cm]{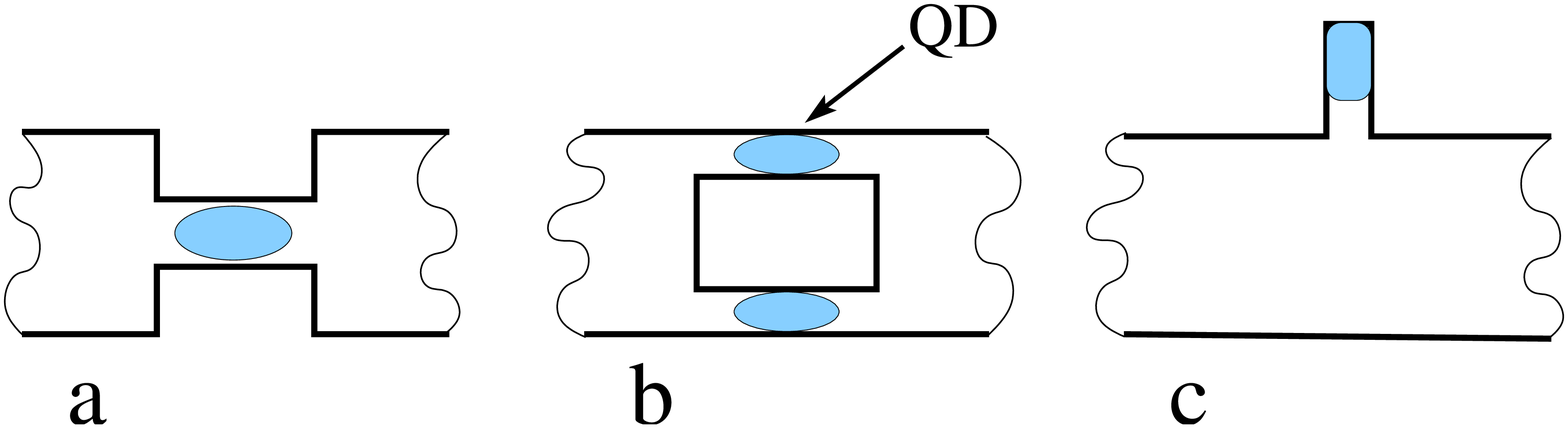} 
\caption{Creation of quantum dots via the charge accumulation in
narrow constrictions. a).~Single QD. b).~Double (parallel) QDs.
c). Side coupled QD. Shaded areas show the lakes of large electron
density. A constriction in biased graphene strip works not as a
Quantum Point Contact, but as a QD.} \vspace{-.3cm}
\end{figure}

A striking consequence of the result, Eq.~(\ref{phistrip}), (see
also Ref.~\cite{Brey07}) for the strip potential is that for $a\ll
b$ the in-plane charge density is inversely proportional to the
strip width. (The linear density of charge $\sigma =\int \rho
(x)dx$ depends only logarithmically on the strip
width,~(\ref{gatecharge}), hence $\rho (x)\sim 1/a$). This offers
a possibility of creating lakes of a large charge density, quantum
dots (QD), by cutting narrow constrictions in the graphene strip.
Examples of such devices are shown in Fig.~6. This semi-mechanical
way of confining electrons (potentials that appear due to the
strip narrowing lead to the longitudinal confinement) may be
complementary to the pure electrical way of fabrication of QDs in
graphene~\cite{SilEf07,Pereira06, DEfetov07}.

Without the electric field doping ($V_{g}=0$) graphene behaves as
a hole metal~\cite{Novoselov04}. The shift of the Fermi energy
away from the Dirac crossing point is attributed to an
unintentional doping of the film by absorbed water. It is
compensated by application of a sufficient lower gate voltage
($\sim 40e{\rm V}$~\cite{Novoselov04}). However, as we have shown,
the gate-induced charging is nonuniform, and it is impossible by
varying $V_g$ to reach the Dirac point simultaneously in the whole
sample. This charge distribution evolves differently for $V_g$
below and above the value corresponding to the strip minimum
conductivity (for example, the local crossing of the Fermi energy
by the Dirac point is shifted towards the center or the boundary
of the strip). This may explain the I-V characteristics assymetry
in graphene (observed e.g. in Ref.~\cite{VanDerSypen07}).

In conclusion, in this paper we predict and describe the
macroscopic charge accumulation along the boundaries of graphene
strips, made of experimentally used mechanically exfoliated films,
for moderate ($\lesssim 10{\rm V}$) lower gate voltages.
Information about the local Fermi momentum and charge density
$p_F\propto \sqrt{n_e}$ may be extracted from the STM measurements
of the density of states in graphene~\cite{Kim07Yakoby07}. The
average charge density $\langle \rho\rangle =\sigma/2a$ for given
gate voltage also strongly increases for narrow strips ($\lesssim
0.5\mu m$) as described by Eq.~(\ref{gatecharge}). Transport in
graphene would be especially sensitive to the predicted charge
accumulation in the experiments~\cite{Novosel05,Zhang05} in the QH
regime, where the two kinds of edge
states~\cite{Chklovskii92,Halperin82,Chamon94} should coexist in
the same sample. Experimental setup capable to measure currents
carried by different edge states is suggested (Fig.~5).

This work was supported by the SFB TR 12. Discussions with
A.F.~Volkov and M.V.~Fistul are greatly appreciated.


\begin{thebibliography}{99}
\bibitem{endnoteQH} One should also avoid cutting sharp angles in
graphene, present e.g. in Ref.~\cite{Novosel05}, where the charge
is electrostatically suppressed.
\end{thebibliography}
\end{document}